\documentstyle[11pt,aaspp4,flushrt]{article}

\lefthead{Stern et al.}
\righthead{A Tidally-Disrupted Star at $z = 3.3$?}

\def\eg{{e.g.,~}}
\def\etal{{et al.~}}

\def\h50min{$h_{50}^{-1}$}
\def\h65{$h_{65}^{-1}$}

\def\pals1{PALS--1}

\def\deg{\ifmmode {^{\circ}}\else {$^\circ$}\fi}

\def\secper{\ifmmode \rlap.{^{s}}\else $\rlap{.}{^{s}} $\fi}

\def\kms{\ifmmode {\rm\,km\,s^{-1}}\else
    ${\rm\,km\,s^{-1}}$\fi}
\def\kmsMpc{\ifmmode {\rm\,km\,s^{-1}\,Mpc^{-1}}\else
    ${\rm\,km\,s^{-1}\,Mpc^{-1}}$\fi}
\def\ergAcm2{\ifmmode {\rm\,ergs\,cm^{-2}\,{\rm \AA}^{-1}}\else
    ${\rm\,ergs\,cm^{-2}\,\AA^{-1}}$\fi}
\def\cm2{\ifmmode {\rm\,cm^{-2}}\else
    ${\rm\,cm^{-2}}$\fi}
\def\ergcm2s{\ifmmode {\rm\,ergs\,cm^{-2}\,s^{-1}}\else
    ${\rm\,ergs\,cm^{-2}\,s^{-1}}$\fi}
\def\cgsdeg2{\ifmmode {\rm\,ergs\,cm^{-2}\,s^{-1}\,deg^{-2}}\else
    ${\rm\,ergs\,cm^{-2}\,s^{-1}\,deg^{-2}}$\fi}
\def\ergsHz{\ifmmode {\rm\,ergs\,s^{-1}\,Hz^{-1}}\else
    ${\rm\,ergs\,s^{-1}\,Hz^{-1}}$\fi}
\def\ergs{\ifmmode {\rm\,ergs\,s^{-1}}\else
    ${\rm\,ergs\,s^{-1}}$\fi}
\def\ergsA{\ifmmode {\rm\,ergs\,s^{-1}\,\AA^{-1}}\else
    ${\rm\,ergs\,s^{-1}\,\AA^{-1}}$\fi}
\def\WHz{\ifmmode {\rm\,W\,Hz^{-1}}\else
    ${\rm\,W\,Hz^{-1}}$\fi}
\def\WHzsr{\ifmmode {\rm\,W\,Hz^{-1}\,sr^{-1}}\else
    ${\rm\,W\,Hz^{-1}\,sr^{-1}}$\fi}
\def\ergscm2Hz{\ifmmode {\rm\,ergs\,cm^{-2}\,s^{-1}\,Hz^{-1}}\else
    ${\rm\,ergs\,cm^{-2}\,s^{-1}\,Hz^{-1}}$\fi}

\def\spose#1{\hbox to 0pt{#1\hss}}
\def\simlt{\mathrel{\spose{\lower 3pt\hbox{$\mathchar"218$}}
     \raise 2.0pt\hbox{$\mathchar"13C$}}}
\def\simgt{\mathrel{\spose{\lower 3pt\hbox{$\mathchar"218$}}
     \raise 2.0pt\hbox{$\mathchar"13E$}}}

\def\plotfiddle#1#2#3#4#5#6#7{\centering \leavevmode
\vbox to#2{\rule{0pt}{#2}}
\includegraphics{#1}}

\received{2004 February 25}
\begin{document}

\title{Discovery of a Transient $U$-band Dropout in a Lyman-Break Survey: \\
A Tidally-Disrupted Star at z = 3.3?\altaffilmark{1,2}}

\author{Daniel~Stern\altaffilmark{3},
P.~G.~van Dokkum\altaffilmark{4,5},
Peter~Nugent\altaffilmark{6},
D.~J.~Sand\altaffilmark{4},
R.~S.~Ellis\altaffilmark{4},
Mark~Sullivan\altaffilmark{9},
J.~S.~Bloom\altaffilmark{10},
D.~A.~Frail\altaffilmark{4,7},
J.-P.~Kneib\altaffilmark{8},
L.\,V.~E.~Koopmans\altaffilmark{4,11},
and Tommaso~Treu\altaffilmark{4,5,12}}

\altaffiltext{1}{Based on observations with the NASA/ESA {\em Hubble
Space Telescope}, obtained at the Space Telescope Science Institute,
which is operated by AURA, Inc., under NASA contract NAS 5--26555.}

\altaffiltext{2} {Based on observations obtained at the W.\ M.\ Keck
Observatory, which is operated jointly by the California Institute of
Technology and the University of California.}

\altaffiltext{3}{Jet Propulsion Laboratory, California Institute of
Technology, Mail Stop 169-506, Pasadena, CA 91109}

\altaffiltext{4}{California Institute of Technology, MS105-24,
Pasadena, CA 91125}

\altaffiltext{5}{Hubble Fellow}

\altaffiltext{6}{Lawrence Berkeley National Laboratory, M.S. 50-F, 1
Cyclotron Road, Berkeley, CA 94720}

\altaffiltext{7}{National Radio Astronomy Observatory, P.O. Box O,
Socorro, NM 87801}

\altaffiltext{8}{Laboratoire d'Astrophysique, Observatoire
Midi-Pyr\'en\'ees, 14 Av.\ E.\ Belin, 31400 Toulouse, France}

\altaffiltext{9}{Physics Department, University of Durham, South Road,
Durham DH 1 3LE, UK}

\altaffiltext{10}{Harvard-Smithsonian Center for Astrophysics, MS 20,
60 Garden Street, Cambridge, MA 02138}

\altaffiltext{11}{Kapteyn Astronomical Institute, P.O. Box 800, 9700AV
Groningen, The Netherlands}

\altaffiltext{12}{Department of Physics and Astronomy, University of
California at Los Angeles, Los Angeles, CA 90095}

\begin{abstract}

We report the discovery of a transient source in the central regions of
galaxy cluster Abell~267. The object, which we call ``\pals1'', was found
in a survey aimed at identifying highly-magnified Lyman-break galaxies
in the fields of intervening rich clusters. At discovery, the source had
$U_n>24.7$ ($2\sigma$; AB), $g=21.96 \pm 0.12$, and very blue $g-r$ and
$r-i$ colors; i.e., \pals1\ was a ``$U$-band drop-out'', characteristic
of star-forming galaxies and quasars at $z\sim 3$.  However, three
months later the source had faded by more than three magnitudes.  Further
observations showed a continued decline in luminosity, to $R>26.4$ seven
months after discovery.  Though the apparent brightness is suggestive of
a supernova at roughly the cluster redshift, we show that the photometry
and light curve argue against any {\em known} type of supernova at
any redshift.  The spectral energy distribution and location near
the center of a galaxy cluster are consistent with the hypothesis that
\pals1\ is a gravitationally-lensed transient at $z\approx 3.3$.  If this
interpretation is correct, the source is magnified by a factor of 4--7
and two counterimages are predicted. Our lens model predicts time delays
between the three images of 1--10 years and that we have witnessed the
final occurrence of the transient.  The intense luminosity ($M_{\rm AB}
\sim -23.5$ after correcting for lensing) and blue UV continuum (implying
$T \simgt$ 50,000 K) argue the source may have been a flare resulting from
the tidal disruption of a star by a $10^{6-8}$\,$M_{\odot}$ black hole.
Regardless of its physical nature, \pals1\ highlights the importance of
monitoring regions of high magnification in galaxy clusters for distant,
time-varying phenomena.

\end{abstract}


\section{Introduction}

The past several years have witnessed dramatic success in our ability to
directly observe the high-redshift ($z \simgt 3$) Universe.  Systematic
photometric selection and narrow-band imaging surveys have identified
several hundred high-redshift galaxies \markcite{Steidel:96b, Stern:99e,
Rhoads:01, Hu:02, Dickinson:04}(\eg Steidel {et~al.} 1996; Stern \&
Spinrad 1999; Rhoads {et~al.} 2001; Hu {et~al.} 2002, Dickinson {et~al.}
2004).  At $z \sim 3 - 4$, these studies have allowed us to directly
measure ensemble properties of the population, such as its luminosity
distribution, color distribution, size distribution, and implied cosmic
star formation rate \markcite{Steidel:99, Shapley:01, Giavalisco:04,
Ferguson:04, Hu:04}(\eg Steidel {et~al.} 1999; Shapley {et~al.} 2001;
Giavalisco {et~al.} 2004; Ferguson {et~al.} 2004, Hu {et~al.} 2004).
At $z \simgt 5$, ambitious programs are underway to extend these studies
to earlier cosmic epoch \markcite{Kodaira:03, Rhoads:04, Kneib:04}(\eg
Kodaira {et~al.} 2003; Rhoads {et~al.} 2004, Kneib {et~al.} 2004).
High-redshift studies are now approaching the dark ages, an epoch when
the Universe was primarily neutral, prior to the reionization imposed
by the first stars and quasars \markcite{Becker:01, Djorgovski:01b}(\eg
Becker {et~al.} 2001; Djorgovski {et~al.} 2001).  A critical void in our
understanding of the earliest phases of galaxy formation has been the
lack of detailed studies of individual sources.  However, with $m_R^* =
24.5$ for the $z \sim 3$ Lyman-break galaxies \markcite{Steidel:99}(LBGs;
Steidel {et~al.} 1999), detailed astrophysical studies are impractical
for the majority of this faint, but important, population.

High-redshift galaxies lensed by rich clusters provide a unique
opportunity for studying in detail the processes which govern galaxy
formation and star formation in the early Universe.  Strong lensing
can increase the brightness of a galaxy by a factor of 10$-$30,
making possible observations which would otherwise be impractical
\markcite{Yee:96, Franx:97, Ellis:01, Kneib:04}(\eg Yee {et~al.}
1996; Franx {et~al.} 1997; Ellis {et~al.} 2001, Kneib {et~al.} 2004).
High-resolution, high signal-to-noise ratio spectroscopy can probe the
ages, kinematics, abundances, mass, and initial mass function of young
protogalaxies \markcite{Pettini:00, Teplitz:00}(\eg Pettini {et~al.}
1999; Teplitz {et~al.} 2000).   Observing normal, high-redshift galaxies
at unusual wavelengths might also become feasible through gravitational
lensing \markcite{Baker:01}(\eg Baker {et~al.} 2001).  In particular,
spectroscopically-confirmed sub-mm galaxies identified by Submillimetre
Common-User Bolometer Array \markcite{Holland:99}(SCUBA; Holland {et~al.}
1999) on the James Clerk Maxwell Telescope (JCMT) have, to date, been
largely restricted to extreme starbursts ($\dot{M} \simgt$ 1000 $M_\odot$
yr$^{-1}$) with some unknown contribution of active galaxies.  It is
thought that the most actively star-forming, high-redshift Lyman-break
systems ($\dot{M} \simgt$ 100 $M_\odot$ yr$^{-1}$) might comprise the
bulk of the sub-mm background \markcite{Adelberger:00}(\eg Adelberger \&
Steidel 2000), just below the {\it unlensed} SCUBA thresholds.  This
hypothesis could be tested with a sample of gravitationally-magnified,
young protogalaxies.

We are undertaking the Palomar Amplified Lyman-break Survey (PALS)
with the aim of enlarging the census of luminous, lensed galaxies at $z
\simgt 3$.  PALS is a $U_ngri$ imaging survey of X-ray selected galaxy
clusters with the 200$''$ Hale telescope at Palomar Observatory. The
well-established color criteria of \markcite{Steidel:95}Steidel, Pettini,
\& Hamilton (1995) and \markcite{Steidel:99}Steidel {et~al.} (1999)
are used to identify candidate magnified, distant galaxies.  Subsequent
spectroscopy with the Keck telescopes is used to confirm the redshifts.

As we show in this manuscript, one of the candidate lensed Lyman-break
galaxies from our survey has faded by more than four magnitudes
since discovery.  This object, which we call ``\pals1'', cannot be
any known type of supernova (SN) at the redshift of the cluster
or neighboring spiral galaxy.  Instead, we suggest it may be the
first optically-selected transient at $z\sim 3$.  High-redshift
optical transients are of considerable interest because of their
links to gamma ray bursts \markcite{Rhoads:01b}(GRBs; \eg Rhoads
2001), SNe searches \markcite{Riess:98, Perlmutter:99}(\eg Riess
{et~al.} 1998; Perlmutter {et~al.} 1999), active galactic nuclei
\markcite{GalYam:02}(\eg {Gal-Yam} {et~al.} 2002b), and accretion of stars
onto black holes \markcite{Rees:88, Komossa:04}(\eg Rees 1988; Komossa
{et~al.} 2004).  Unfortunately, the assembled data is insufficient to
unambiguously describe the physical nature of \pals1.  We argue that
the most likely interpretations are that we have witnessed either a
previously-unidentified, peculiar type~Ia SN, or we have witnessed the
tidal disruption of a star by the dormant, supermassive black hole in the
center of a faint, normal galaxy.  The latter events reveal themselves
by a UV/X-ray flare and were predicted by \markcite{Lidskii:79}Lidskii
\& Ozernoi (1979) and \markcite{Rees:88}Rees (1988).  The {\it ROSAT}
All-Sky Survey identified several plausible soft X-ray candidates
\markcite{Li:02}(see Li, Narayan, \& Menou 2002, and references therein).
We are aware of no robust UV/optical candidates published to date.
The paper is organized as follows.  In \S~2 we discuss the observations.
In \S~3 we explore likely physical interpretations of this unusual
flare event, followed by an estimate of their likelihoods in \S~4.
Our conclusions comprise \S~5.

\section{Observations}

\subsection{Spectral Energy Distribution on UT 2001 July 20}

As part of the PALS survey, the $z=0.23$ cluster Abell~267 was observed
with the Carnegie Observatories Spectoscopic Multislit and Imaging
Camera \markcite{Kells:98}(COSMIC; Kells {et~al.} 1998) on the Palomar
200$''$ telescope on UT 2001 July 20.  Exposure times were 900\,s in
$U_n$ \markcite{Steidel:03}($\lambda_c = 3550$\,\AA; $\Delta\lambda =
600$\,\AA; Steidel {et~al.} 2003), 600\,s in $g$ ($\lambda_{\rm c} =
5000$\,\AA; $\Delta\lambda = 800$\,\AA), 240\,s in $r$ ($\lambda_{\rm
c} = 6550$\,\AA; $\Delta\lambda = 900$\,\AA), and 240\,s in $i$
($\lambda_{\rm c} = 8000$\,\AA; $\Delta\lambda = 1800$\,\AA).
Conditions were photometric, and the seeing was $\approx 1\farcs 6$
in all bands.  Photometric calibration was performed using the Sloan
Digital Sky Survey (SDSS) Early Data Release observations of Abell~267
\markcite{York:00}(York {et~al.} 2000).  Throughout, magnitudes are
referred to the AB system \markcite{Oke:83}(Oke \& Gunn 1983).

\begin{figure}[!t]
\begin{center}
\plotfiddle{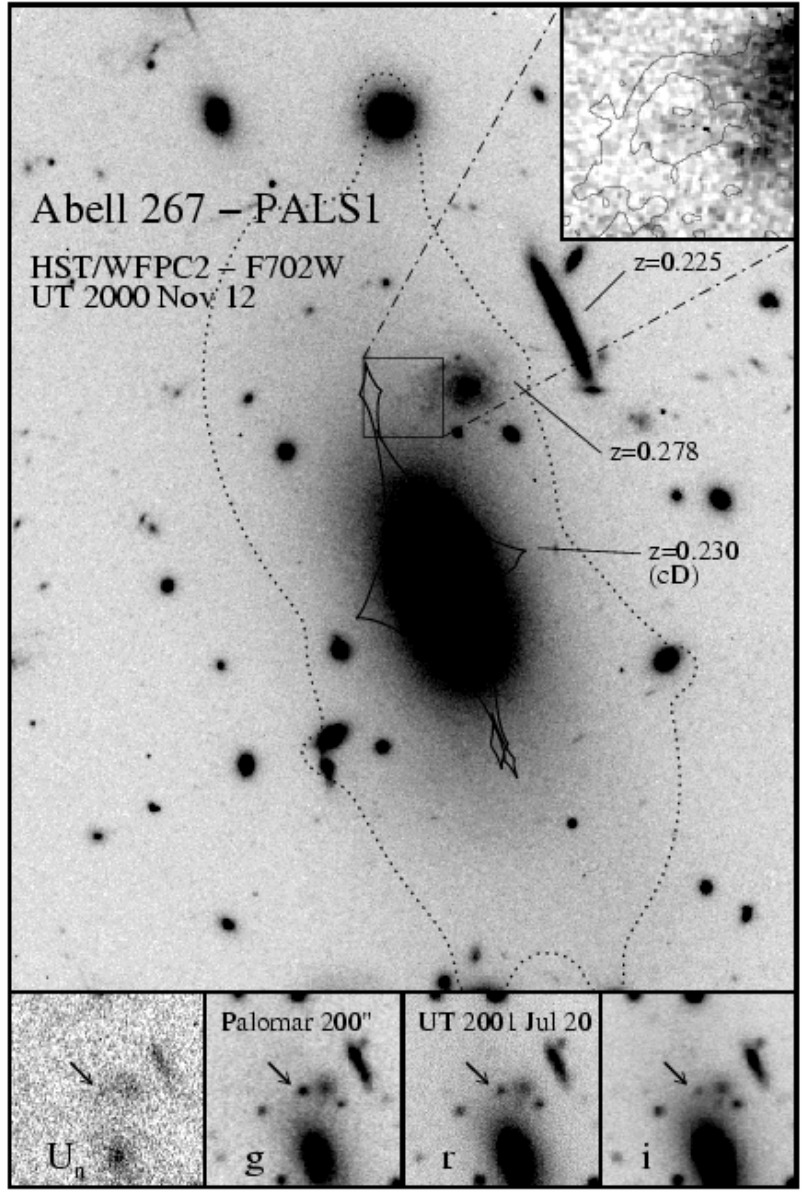}{5.2in}{0}{80}{80}{-275}{-100}
\end{center}

\figcaption{{\bf [SEE f1.gif]}.  Images of \pals1\ in Abell~267.  The bottom panels show the
discovery multi-band imaging of \pals1, obtained on UT 2001 July 20
with the Palomar 200$''$ telescope.  The central panel, subtending
50\arcsec\ $\times$ 62\farcs5, shows the {\it HST}/WFPC2 $R_{F702W}$
image taken on UT 2000 November 12.  North is up and east is to the
left.  The dotted contour shows the critical line for a source at
$z=3.3$, and the solid contour shows the caustic. In the lensing
hypothesis, \pals1\ would be near a cusp in the source plane and give
rise to three images.  The top inset, 5\arcsec\ on a side, shows the
{\it HST} image at the location of \pals1.  The light of the cD galaxy
has been subtracted and contours indicate isophotal contours from the
discovery $r$-band image.  A faint ($R = 27.7$; $3\sigma$) source is
detected, slightly offset from the transient location.  This is likely
associated with either the galaxy or the star-forming region which hosted
the optical transient.\label{image.plot}}
\end{figure}

The survey located an object (see Fig.\ \ref{image.plot}) in the envelope
of the central cD galaxy with the colors of a young galaxy or quasar
at $z\sim 3$.  The object, which we call \pals1, is located at R.A. $=
1^h52^m42.0^s$, Dec. $=1\deg00\arcmin37\farcs5$ (J2000).   As shown in
Fig.\ \ref{sed.plot}, at discovery \pals1\ had a very blue continuum
redward of $\sim 4500$\,\AA, and a strong break between the $U_n$-band
and the $g$-band.  With $U_n>24.7$ ($2 \sigma$), $g=21.96\pm 0.12$, $g-r
= -0.2 \pm 0.2$ and $r-i=-0.4 \pm 0.2$ (see Table\ \ref{sed.table}), the
candidate Lyman-break object was situated in a region of color-color space
far removed from Galactic and low redshift objects \markcite{Steidel:95,
Fan:99}(Steidel {et~al.} 1995; Fan {et~al.} 1999).  The colors of \pals1, in combination with its brightness
and its location near the central cD galaxy in Abell~267, made it a
promising candidate lensed, high-redshift object.  In particular, \pals1\
is approximately 2 magnitudes brighter than $m_*$ of the $z \sim 3$
Lyman-break population \markcite{Steidel:99}(Steidel {et~al.} 1999), comparable to the gravitational
magnification provided by our lensing model of the cluster (\S 3.2.1).

\subsection{Light Curve and Host}
\label{timeevo.sec}

On UT 2001 October 18 we intended to obtain a spectrum of \pals1\ with
the Keck I Low Resolution Imager and Spectrograph \markcite{Oke:95}(LRIS;
Oke {et~al.} 1995), but found that the source was no longer visible on the
guider. On UT 2001 October 21 we imaged the field with the Keck II Echelle
Spectrograph and Imager \markcite{Sheinis:02}(ESI; Sheinis {et~al.} 2002).
The ESI images showed a faint red source at the location of \pals1,
with Spinrad $R_s=25.3 \pm 0.15$ \markcite{Djorgovski:85a}($\lambda_c =
6890$~\AA; $\Delta \lambda = 1370$~\AA; Djorgovski 1985) and Johnson $B =
26.7 \pm 0.2$ ($\lambda_c = 4400$~\AA; $\Delta \lambda = 980$~\AA). We
conclude that the object had faded by $3.2 \pm 0.2$ magnitudes in $R$
over the course of 93 days.  The apparent change in color is difficult
to interpret because the $B$ filter lies blueward of the $g$ filter.
Keck I Near Infrared Camera \markcite{Matthews:94}(NIRC; Matthews \&
Soifer 1994) observations give an upper limit of $K_{\rm AB}>24.7$
($3 \sigma$) on UT 2002 October 31, and radio observations with the Very
Large Array on the same day give a $3\sigma$ upper limit of 0.08\,mJy at
8.5\,GHz{\altaffilmark{13}.  The field was re-observed several times in
the optical during November 2001 -- February 2002. The source continued
to fade over this time interval, to $R > 26.4$ on UT 2002 February 16
(see Table\ \ref{tableLC}).

\altaffiltext{13}{We note that our preferred physical description of
\pals1, a star being tidally disrupted by a supermassive black hole,
is clearly related to variability of an AGN, in the limit of a very
low accretion rate.  Namely, we are witnessing the accretion of a
single star.}

Archival observations of Abell~267 were obtained from the Digitized
Palomar Observatory Sky Survey \markcite{Djorgovski:99}(DPOSS; Djorgovski {et~al.} 1999), the
Near Earth Asteroid Tracking survey \markcite{Pravdo:99}(NEAT; Pravdo {et~al.} 1999), the
archive of the Isaac Newton Group (ING) of telescopes, the Canadian
Astronomy Data Center \markcite{Dahle:02}(CADC; Dahle {et~al.} 2002), and the {\em Hubble
Space Telescope} ({\em HST}) Archive. A constant $f_{\nu}$ long-ward of
$\approx 5000$~\AA\ (approximate for \pals1\ around the time of
discovery) was assumed in cases where $R$-band observations were not
available.

\begin{figure}[!t]
\begin{center}
\plotfiddle{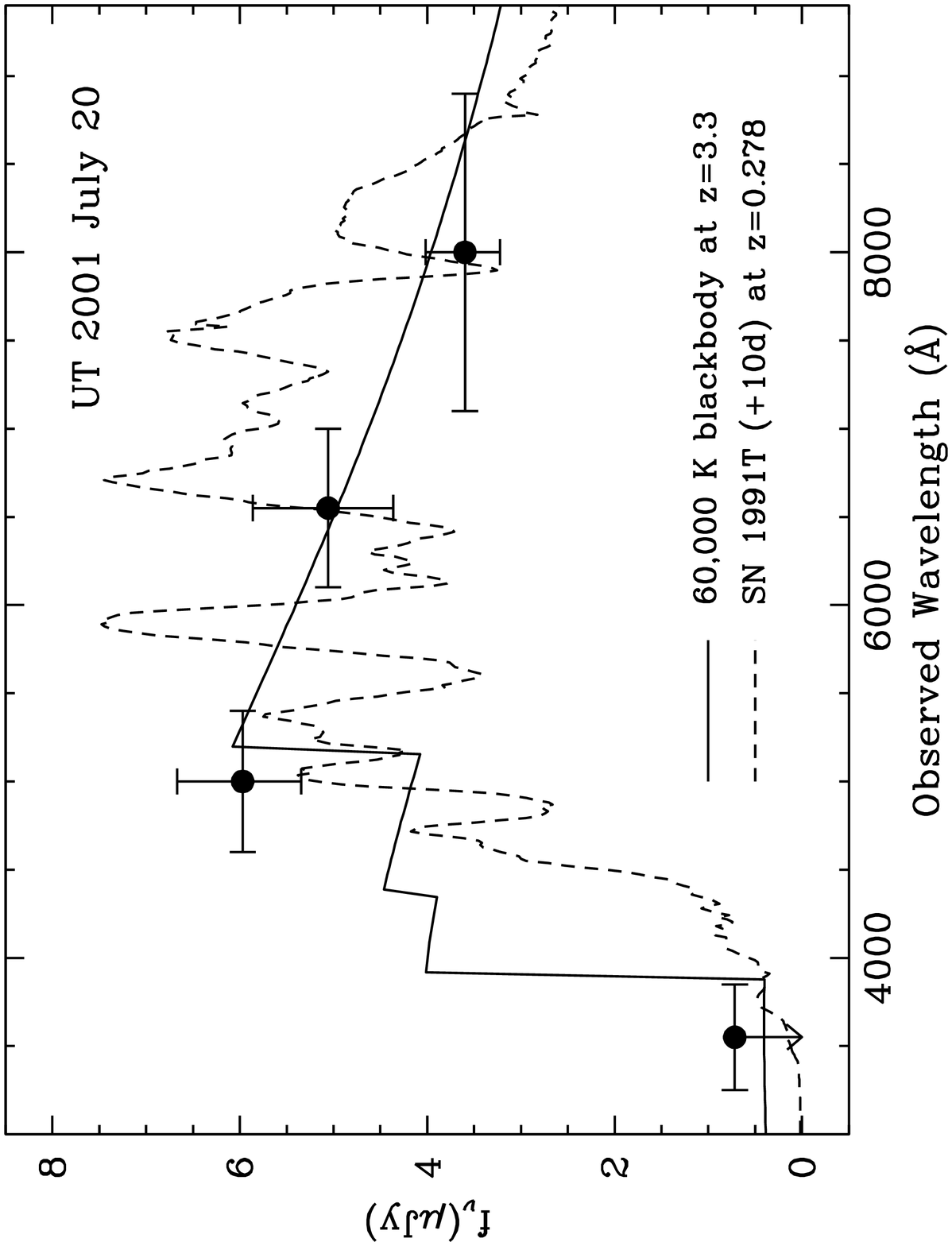}{3.8in}{-90}{55}{55}{-235}{335}
\end{center}

\figcaption{Spectral energy distribution of \pals1\ on UT 2001 July 20
(solid circles).  The arrow indicates a $3\sigma$ upper limit for
the $U_n$-band photometry.  Spectra for possible transient object
interpretations are overplotted.  The solid line shows a 60,000~K black
body spectrum redshifted to $z=3.3$ and subject to hydrogen absorption
from interstellar and intergalactic hydrogen.  The dashed line shows
the scaled spectrum of the type~Ia SN~1991T at 10 days after peak
brightness, dereddened by $E(B-V) = 0.22$ and redshifted to $z = 0.278$.
The spectrum is a combination of {\em IUE} UV data (Jeffery \etal 1992)
and Cerra Tololo optical data (Phillips \etal 1992).  For the SN model,
our best-fit match to the colors is a $z = 0.278$ 1991T-like type~Ia SN
7.25 days after peak brightness (in the observers' frame; \S 3.1).
\label{sed.plot}}
\end{figure}

No further detections were found, except for two cases.  Slightly offset
 from the location of \pals1\
(0\farcs2 $\pm$ 0\farcs1 in each coordinate), a marginally extended object
is detected at $3\sigma$ above the local background in an $R_{F702W}$
($\lambda_c = 6818$~\AA; $\Delta \lambda = 1385$~\AA) {\em pre-discovery}
image taken on UT 2000 November 12 with the {\em HST} Wide Field Planetary
Camera 2 \markcite{Trauger:94}(WFPC2; Trauger, Ballester, \& Burrows 1994).  The {\em HST} image, shown in
Fig.\ \ref{image.plot}, was obtained in a survey aimed at measuring
the mass distribution of a sample of X-ray clusters at $z\sim 0.2$
\markcite{Smith:01}(see Smith {et~al.} 2001).  The counterpart to \pals1\ is very faint at $R
= 27.7^{+0.4}_{-0.3}$ and is likely associated with either the galaxy
or star-forming region which hosted the optical transient.  The other
detection is in data taken on UT 2001 July 25, i.e., 5 days after the
Palomar discovery images.  During a spectroscopic observation with ESI
of an unrelated cluster galaxy, an unfiltered snapshot was taken with
the guider camera.  An approximate photometric calibration was obtained
from blue stars in the guider field.

The light curve of \pals1\ constructed from these various observations
is shown in Fig.\ \ref{evo.plot}. Shallow (19--20 mag) limits from NEAT
are not shown. The decay after JD\,52110.5 can be approximated by an
exponential of the form $f_{\nu} \propto e^{\alpha t}$, with $\alpha
= -0.033 \pm 0.001$ and $t$ measured in days.  We note that since the
Palomar observations were not taken concurrently, the apparent $U_n$-band
drop could be due to rapid variability.  However, given the long-timescale
secular fading seen in the $R$-band, we consider short-term variability
an unlikely explanation for the ultraviolet drop.

\begin{figure}[!t]
\begin{center}
\plotfiddle{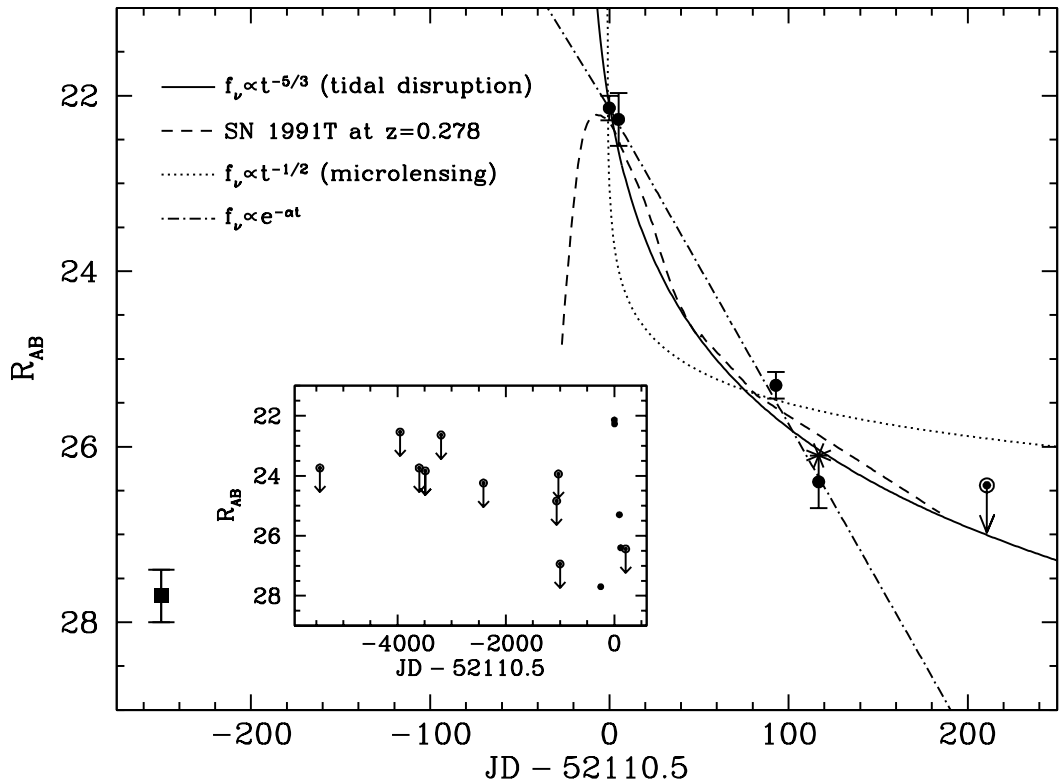}{3.8in}{-90}{55}{55}{-235}{335}
\end{center}

\figcaption{Light curve in the $R$ band; arrows indicate $2\sigma$ upper
limits.  The data point at $\Delta {\rm JD} = -250$ comes from the {\it
HST} image and is assumed associated with the host galaxy.  The data
point at $\Delta {\rm JD} = 117$ was corrected for the luminosity of
the underlying host (see Fig.\ \ref{image.plot}); the asterisk shows
the uncorrected measurement.  Lines show temporal evolution of various
models, all fitted to the data after JD\,52110.5.  The solid line shows
a power-law with index $-5/3$, as expected for the flare generated by a
tidally-disrupted star (\S~3.2).  The dashed line shows our best-fit SN
model (\S~3.1), an underluminous 1991T-like type~Ia SN at $z = 0.278$
with maximum brightness occurring 7.25~days prior to the discovery images.
The dotted line shows a power-law with index $-1/2$ (dotted), as expected
for a microlensing event.  The dot-dashed line shows an exponential fit.\label{evo.plot}}
\end{figure}

\section{The Redshift of \pals1}
\subsection{A Supernova at $z \approx 0.25$?}

\pals1\ is located $4\farcs 5$ east of a bright spiral galaxy (see
Fig.~\ref{image.plot}) and we first consider the hypothesis that the
transient is a SN in the outskirts of this galaxy. In this interpretation
the object detected in the {\em HST} image is a star forming region at
the same redshift as the spiral galaxy. A 2400\,s long slit spectrum
containing both \pals1\ and the spiral galaxy was obtained on UT 2001
October 22 with ESI on Keck II. The redshift of the spiral galaxy is
$z=0.278$, as determined from H$\alpha$, [N\,{\sc ii}], and [S\,{\sc ii}]
emission lines.  We now demonstrate that no {\it known} SNe are able to
simultaneously match the colors and brightness of \pals1.

The rest-frame rate of decay for $z=0.28$ would be approximately
$0.046$\,mag\,day$^{-1}$, and the brightness at discovery would be $M_g
\sim -18$ after a 0.5 mag correction for lensing. These values are in the
range expected for type~I and type~IIL SNe \markcite{Filippenko:97}(\eg
Filippenko 1997).  However, the colors of \pals1\ at the time of discovery
are incompatible with most types of SNe at this redshift range.  Type~II
SNe at $z \sim 0.25$ only become red in $U_n - g$ at late stages ($\simgt
100$ days), while \pals1 is red in $U_n - g$ near maximum brightness.
Furthermore, when $z \sim 0.25$ type~II SNe exhibit red $U_n - g$ colors,
they are red across the optical window, which does not match the discovery
spectral energy distribution (SED) of \pals1.  Finally, type~II SNe do
not fade fast enough to match our 4 magnitudes of dimming in 100 days.
We conclude that \pals1\ is unlikely to be a type~II SN.

Similarly, type~Ib/c SNe can be ruled out since the 4 magnitudes of
fading in $\sim 100$ days would require that our discovery imaging
occurred right at maximum, at which time the SED is problematic.
\markcite{Poznanski:02}Poznanski {et~al.} (2002) predict that any
normal type~I SN at $z \sim 0.28$ has $g - r > 0.4$ at all epochs,
compared to the observed $g - r = -0.2 \pm 0.2$ for \pals1\ at discovery
(Fig.~\ref{SNe.plot}).  Similar considerations lead us to conclude that
any normal form of type~I or type~II SN at $z < 1$ can be ruled out as
an explanation of \pals1.

\begin{figure}[!t]
\begin{center}
\plotfiddle{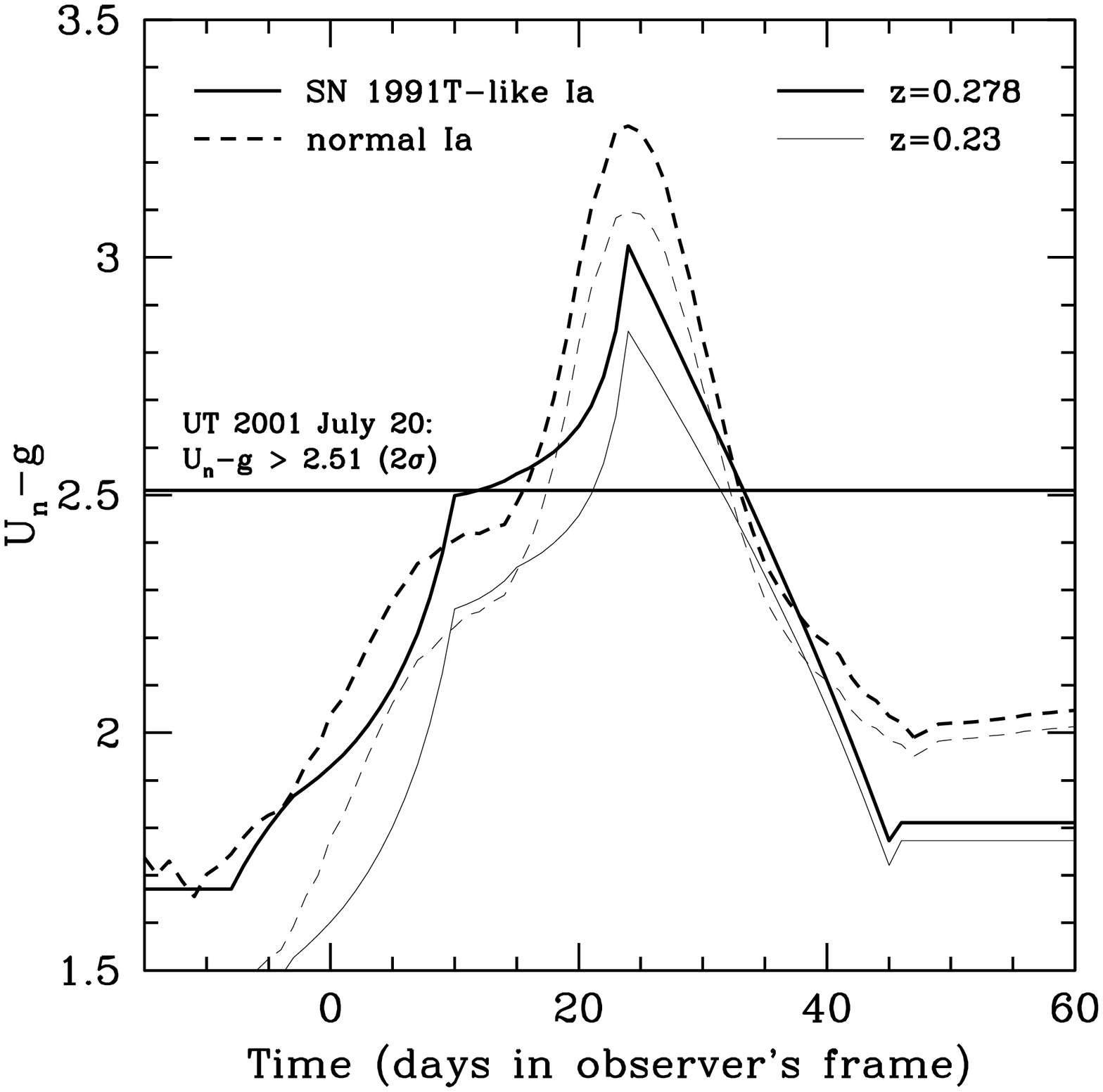}{2.2in}{0}{40}{40}{-250}{-95}
\plotfiddle{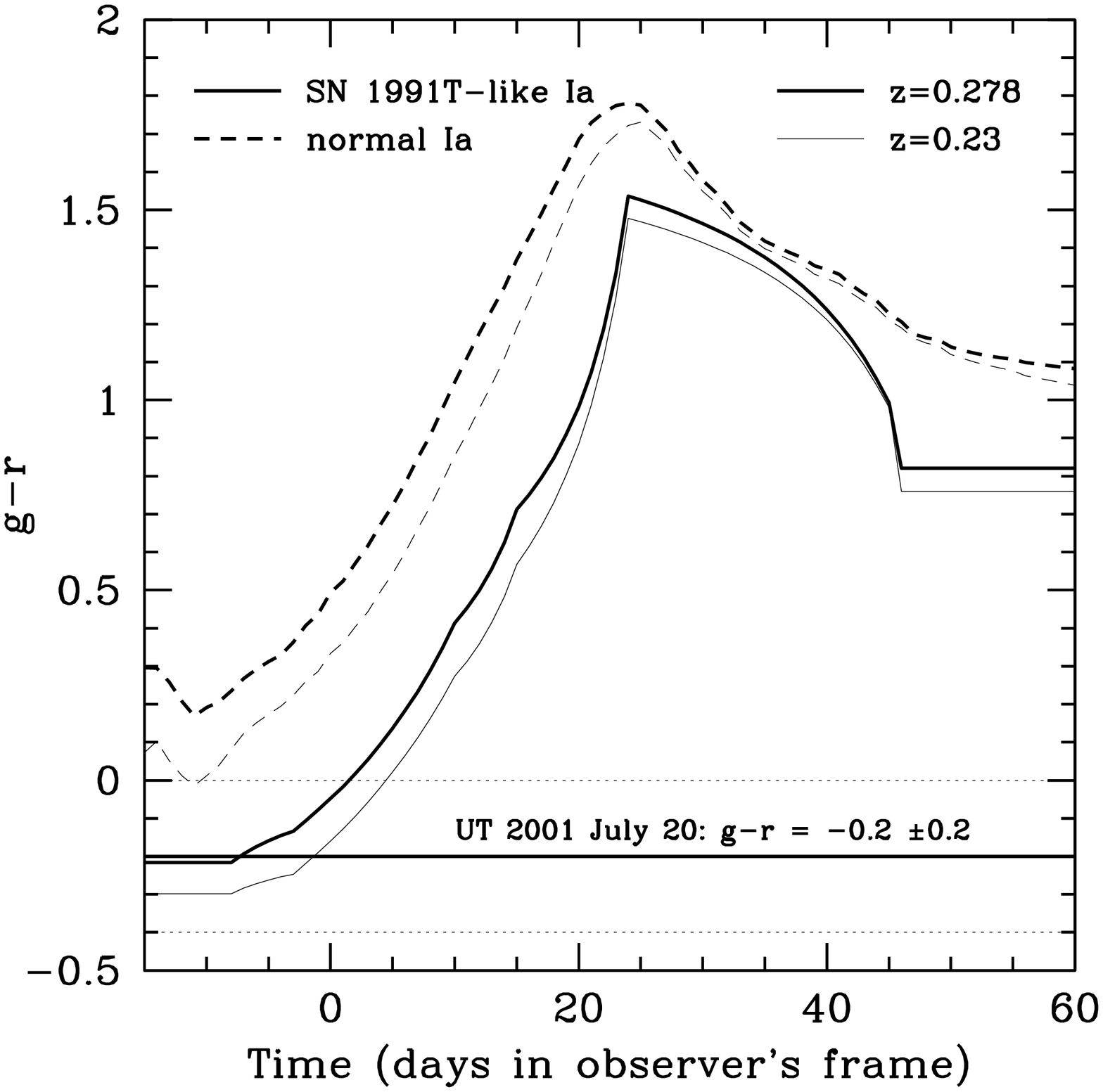}{0.0in}{0}{40}{40}{0}{-65}
\end{center}

\figcaption{Color evolution of type~Ia SNe at $z = 0.278$ (darker lines;
the redshift of the nearby spiral galaxy) and $z = 0.230$ (thinner lines;
the redshift of galaxy cluster Abell~267).  Dashed lines illustrate the
color evolution for normal type~Ia SNe, while solid lines show color
evolution for 1991T-like type~Ia SNe.  The red $U_n - g$ color of PALS-1
at discovery is matched by type~Ia SNe at $z \sim 0.25$ for observed times
approximately $10 - 30$ days after maximum.  However, the blue $g - r$
color of PALS-1 at discovery is difficult to self-consistently match.
A normal type~Ia SN should have been a least one magnitude redder in
$g - r$ at this epoch.  A 1991T-like type~Ia SNe is consistent with
the observed colors at $\sim$ 10 days after peak brightness, but has
significant problems matching the observed brightness.\label{SNe.plot}}
\end{figure}

One peculiar form of type~Ia SN, resembling SN~1991T
\markcite{Filippenko:92}(Filippenko {et~al.} 1992), does, however, have a
marginally acceptable fit to the colors of \pals1\ (Fig.~\ref{SNe.plot}).
Such SNe represent 5--20\% of the type~Ia population \markcite{Branch:93,
Li:01}(Branch, Fisher, \& Nugent 1993; Li {et~al.} 2001a).  They are
characterized by dusty environments and are typically $\sim 0.4$~mag
overluminous in $V$ \markcite{Nugent:95}(Nugent {et~al.} 1995).
A $\chi^2$ fit to the SED and light curve of \pals1, assuming $z = 0.278$,
finds a peak brightness occurring at JD~52103.25 (\eg 7.25 days prior to
our discovery imaging) with no extinction and a peak, corrected magnitude
$R = 22.22$.  Fig.~\ref{sed.plot} illustrates our discovery photometry
with the observed spectrum of SN~1991T at 10~days after peak brightness
\markcite{Jeffery:92, Phillips:92}(Jeffery {et~al.} 1992; Phillips
{et~al.} 1992).  Adopting $z = 0.23$, the cluster redshift, degrades
the fit.  For these analyses, we applied a 0.23~mag color term correction
to the $U_n$ magnitude limit to account for the differences between the
$U_n$ and SDSS $u'$ filter convolutions with blue stars (which were used
to calibrate the Palomar images; \S~2.1) and early-time 1991T-like SNe at
$z = 0.278$.  The correction is adopted in Fig.~\ref{SNe.plot} and implies
that, at discovery, \pals1\ had $U_n > 24.47$ for the SN interpretation.

A normal type~Ia SN at $z = 0.278$ would have a peak apparent magnitude
of $R = 21.25$.  SN~1991T-like type~Ia SN are typically a few tenths of
a magnitude overluminous, implying an expected apparent magnitude
of $R = 21.0$ at peak, or $R = 21.1$ at $+7.25$~days.  For $z = 0.278$,
our lens model (see \S 3.2.1) implies \pals1\ has been magnified by
$\approx 0.5$ mag, implying that the {\em expected} apparent magnitude of
\pals1\ in our discovery images would have been $R = 20.6$.  This is
1.5~magnitudes brighter than what was actually seen.  This brightness
offset is quite worrisome.   Though 1991T-like SNe are always associated
with some extinction, attributing the brightness offset to dust extinction
will redden the expected SED and aggravate the color fit to the discovery
photometry.  Additionally, 1991T-like SNe typically occur in strong
star-forming regions, closer to the host galaxy nuclei than \pals1\
is observed relative to the presumed $z = 0.278$ host spiral galaxy.
We conclude that \pals1\ can not be a 1991T-like type~Ia SN event.

Finally, since a SN interpretation for \pals1\ requires an underluminous
type~Ia event, we consider known underluminous type~Ia SNe.  The best
studied of this class are 1991bg-like type~Ia SNe.  However, since
Ti absorption causes such SNe to be red in $g - r$ at all epochs,
we conclude that \pals1\ can not be such an event.  \markcite{Li:03}Li {et~al.} (2003b)
recently reported on SN~2002cx, a peculiar type~Ia SN, unlike any known
prior SNe.  SN~2002cx exhibits a SN~1991T-like premaximum spectrum, but a
SN~1991bg-like luminosity, suggesting it could be a good match to \pals1.
However, SN~2002cx reddens very rapidly after maximum, making a poor
fit to the discovery photometry of \pals1, which is expected to coincide
with approximately a week after peak brightness.  We conclude that all
{\em known} forms of SNe can be ruled out as an explanation of \pals1,
though the possibility exists that a new form of SN could be identified
which would match the colors, lightcurve, and brightness of \pals1.

\subsection{A Gravitationally-Lensed Transient at $z \approx 3.3$?}

\subsubsection{Lens Model}

Another natural explanation for the $U$-band ``dropout'' signature seen
in \pals1\ on UT 2001 July 20 is absorption by the interstellar medium
(ISM) and intergalactic medium (IGM) of a UV bright source at $z\sim
3$.  The SED can then be fit equally well by a power law with index
$\beta_{\nu}= 1.0 \pm 0.4$ or a thermal spectrum. As an example, the
dashed line in Fig.\ \ref{sed.plot} shows a 60,000~K thermal spectrum at
the best-fit redshift $z=3.3$, including the effects of the ISM and
IGM. Temperatures $\simgt$ 50,000~K and redshifts $3.1\simlt
z\simlt 3.5$ provide good fits to the photometric data.

If the Lyman-break interpretation is correct, the source is strongly
lensed by Abell 267. The mass distribution of the cluster was modeled
following the techniques described in \markcite{Kneib:96}Kneib {et~al.}
(1996) and is discussed in full in G.\ Smith et al.\ (in preparation). The
absolute mass distribution is somewhat uncertain because no arc redshifts
have yet been measured in this cluster.  Accordingly, we consider two
mass models spanning the likely range.  Assuming \pals1\ is in the
redshift range $3.1<z<3.5$, consistent with the Palomar photometry,
the amplification at the location of \pals1\ is a factor 4--7, and two
counterimages are predicted, magnified by factors of 3.5--5.  Although
several candidate counterimages can be identified in the {\em HST}
image, deeper multi-color data are required to unambiguously identify
the counterimages.

The correspondence between the $z \sim 3$ Lyman-break colors of
\pals1\ and its location near the $z \sim 3$ critical line of our
Abell~267 lens model argues strongly that the object may indeed be at
this high redshift.  Although highly-magnified images of distant
objects can be found out to $\sim 30''$ from the cD galaxy, they should
lie in restricted regions typically close to the critical line.
Considering the area within $30\arcsec$ of the cD, the probability that
\pals1\ would by chance be located at a position where highly-magnified
$z\sim 3$ objects can be expected is only $\sim 0.02$.

The lens model predicts differing light travel times for the lens
images, and implies that \pals1\ represents the 
the final occurrence
of the transient. The absolute time delays between the images are not well
constrained, but are in the range 1--10 years.  We carefully examined
the archival data discussed in \S~\ref{timeevo.sec} to look for a
previous occurrence of the transient in a counterimage.  As might have
been expected from the sparse sampling in the time domain, no
convincing cases of previous occurrences were found.

\subsubsection{A Tidally-Disrupted Star?}

The potential description of \pals1\ as a SN at the redshift of
nearby galaxies is thoroughly discussed in \S 3.1.  We now speculate on
the physical nature of \pals1, assuming it is a lensed transient at $z
\sim 3$.  The luminosity in the far UV at the time of discovery
($M_{\rm AB} \sim -23.5$ after correcting for lensing) is orders of
magnitude larger than expected from any SN.  GRBs have typical
spectral indices $-1.5 \simlt \beta_{\nu} \simlt -0.5$ in the
optical \markcite{Rhoads:01b}(Rhoads 2001), much redder than the observed index
$\beta_{\nu} = 1.0 \pm 0.4$.  Variability of an Active Galactic Nucleus
(AGN) is possible but not likely, given the lack of detections in
numerous archival observations, the large amplitude of the observed
variation as compared to typical ``violently variable'' AGN
\markcite{Pollock:79, GalYam:02, Geha:03}(\eg Pollock {et~al.} 1979; {Gal-Yam} {et~al.} 2002b; Geha {et~al.} 2003), and the non-detection at
8.5\,GHz. Micro-lensing of a radio-quiet, optically-faint
AGN by a compact object in the cD halo \markcite{Walker:95}(Walker \& Ireland 1995) is highly
unlikely, because the luminosity does not decay as $L \propto t^{-1/2}$
at late times and it would require that the event was observed within
hours of maximum brightness (see Fig.\ \ref{evo.plot}).

An intriguing possibility, consistent with all observations to date, is
that \pals1\ was a flare resulting from the tidal disruption of a star
by a $10^{6-8}$\,$M_{\odot}$ black hole \markcite{Rees:88}(Rees 1988). Such events
are predicted to be very luminous and to give rise to extremely blue
spectra, corresponding to effective temperatures $> 10^4$\,K
\markcite{Ulmer:99}(\eg Ulmer 1999). Both the UV luminosity and the blackbody
temperature of \pals1\ are within the range expected for these events.
The time scale can vary significantly, ranging from weeks to years. In
the ``fallback stage'' \markcite{Rees:88}(Rees 1988) the luminosity is expected to
decay as a power-law with exponent $-5/3$. As shown by the dashed line in
Fig.\ \ref{evo.plot}, the observed decay of \pals1\ is consistent with
this power-law index, provided that the flare was observed $\approx 15$
days ($\approx 3$ days in the rest frame) after the initial tidal
disruption.

The slight offset between the observed flare and the presumed faint host
identified in the {\it HST} images is slightly worrisome for the stellar
disruption scenario.  However, for $z \sim 3.3$, the {\it HST} image
would trace rest frame UV ($\approx 1650$ \AA) light, which generally
does not trace the mass of a star-forming galaxy, and thus would not
necessarily coincide with the galaxy nucleus.

Detection of UV flares is of great interest because it may be the
only way to detect massive black holes in non-active galaxies at $z
\simgt 3$. The time domain is just starting to be explored at faint
levels \markcite{DellAntonio:01}(\eg {Dell'Antonio} {et~al.} 2001), and we may expect to find
unlensed examples of this type of transient.  Indeed, \markcite{Genzel:03}Genzel {et~al.} (2003)
and \markcite{Ghez:03}Ghez {et~al.} (2003) recently reported near-infrared flares from the
supermassive black hole at the center of the Milky Way, possibly
associated with the accretion of comet- or small asteroid-sized
bodies.  The photometric properties of flares are thought to depend on
the mass of the black hole \markcite{Ulmer:99}(Ulmer 1999), and with sufficiently
large samples it may be possible to extend studies of the relation
between massive black holes and their host galaxies
\markcite{Ferrarese:00}(\eg Ferrarese \& Merrit 2000) to very high redshift.

\section{Event Rates}

Though we have shown that \pals1\ is unlike any {\em known} type of SN,
the possibility exists that we have discovered a new, rare type of SN,
similar to a 1.5 magnitude underluminous 1991T-like type~Ia SN.  In the
past four years alone, three ``unique'', never-seen-before, nearby type-Ia
SNe have been discovered:  SN~2000cx \markcite{Li:01b}(Li {et~al.}
2001b), SN~2002cx \markcite{Li:03}(Li {et~al.} 2003b), and SN~2002ic
\markcite{Deng:04}(Deng {et~al.} 2004).  Underluminous SNe would likely
be under-represented in flux-limited surveys.  Alternatively, \pals1\
may be associated with the optical flare caused by a tidally-disrupted
star, also a never-seen-before phenomenon.  Subject to the uncertainties
inherent in estimating the rates of such singular events, we next
compare the relative likelihoods of a peculiar SN at $z \approx 0.25$
and a gravitationally-lensed, tidally-disrupted star at $z \approx 3.3$.
Again, we emphasize that neither phenomena have been observed previously,
so the following estimates are subject to substantial uncertainties.

\subsection{Peculiar Type~Ia Supernovae at $z \approx 0.25$}

Based on the color and apparent brightness of \pals1\ at discovery, a
SN identification would likely be near the cluster redshift.  Since the
light curve of \pals1\ argues against it being associated with a type~II
SN (\S~3.1), we begin our estimate with the type~Ia SN rate in galaxy
clusters; field galaxies at similar redshifts are unlikely to increase
this number by more than a small factor.  \markcite{GalYam:02b}{Gal-Yam},
Maoz, \&  Sharon (2002a) recently measured the type~Ia SN event rate for
galaxy clusters, finding $0.20^{+0.84}_{-0.19}~ h_{50}^2$ type~Ia SNe
century$^{-1}$ per $10^{10}~L_{B \odot}$ for $ 0.18 \leq z \leq 0.37$
galaxy clusters.  Typical galaxy clusters have $2.5 \times 10^{12}~
h_{50}^{-2}~ L_{B \odot}$.  As argued in \S~3.1, for this scenario
\pals1\ is likely related to a 1991T-like type~Ia SN.  Such SNe represent
5$-$20\%\ of the type~Ia population \markcite{Branch:93, Li:01}(Branch
{et~al.} 1993; Li {et~al.} 2001a).  Perhaps as much as 5\%\ of the
1991T population could match our underluminosity requirement, {\it
i.e.,} $\simlt$ 0.5\%\ of the type~Ia SN population are potentially
peculiar and could match \pals1.  This is the largest uncertainty in
our estimate; current and future ambitious SNe surveys will refine, and
possibly measure, this number.  A SN at $z \approx 0.25$ would have been
visible to our survey for approximately two months, though the red $U -
g$ and blue $g - r$ colors which would have caused us to notice such a SN
are only present for $\approx$ 12 days.  Folding these factors together,
we estimate a $\simlt (4 - 400) \times 10^{-6}$ probability that \pals1\
is an as-yet-unknown, peculiar type~Ia SN in Abell~267 at $z = 0.23$.
We emphasize that the photometry analysis found a better match for a SN
in the background $z = 0.278$ spiral galaxy (\eg Fig.~\ref{SNe.plot})
rather than in the cluster, implying that this estimate is potentially
several orders of magnitude optimistic.

\subsection{Gravitationally-Lensed, Tidally-Disrupted Stars at $z \approx 3.3$}

According to the hypothesis that \pals1\ is a $z \approx 3.3$
tidally-disrupted star gravitationally-lensed by Abell~267, the $R =
27.7$ source identified in our {\it HST} image is the host galaxy.
Our lens model (\S~3.2.1) implies $1.5 - 2.1$ magnitudes of gravitational
magnification.  We first calculate the likelihood that an intrinsically $R
\approx 29.5$ LBG at $z \sim 3$ has been highly magnified by the cluster,
and we then use the theoretical stellar disruption rate calculated by
\markcite{Magorrian:99}Magorrian \& Tremaine (1999) to estimate the
probability of the tidal disruption scenario.

The $z \sim 3$ LBG luminosity function from \markcite{Steidel:99}Steidel
{et~al.} (1999) implies a surface density of 30.4 LBGs arcmin$^{-1}$ for
$29 < R < 30$.  Lensing magnification by a factor of five dilutes the
surface density by the same factor.  Since only 2\%\ of the area with
30\arcsec\ of the cluster cD galaxy is highly magnified, we estimate
$\approx$ 0.1 faint ($R \approx 29.5$), $z \sim 3$ LBGs are highly
magnified by Abell~267.

Based on dynamical models of real galaxies with supermassive, central
black holes, \markcite{Magorrian:99}Magorrian \& Tremaine (1999)
calculate approximately one star is tidally disrupted per $10^4$ yr
for faint ($L \simlt 10^{10} L_\odot$) galaxies.  The event rate is
approximately flat for $10^{8.5} < L / L_\odot < 10^{10}$.  More massive
galaxies have more massive black holes which swallow dwarf stars whole
and hence do not emit flares.  This, combined with the flatter central
density profiles in such galaxies, causes the tidal disruption rate to
decrease for more massive galaxies.

For the \markcite{Steidel:99}Steidel {et~al.} (1999) rest frame
UV $z \sim 3$ LBG luminosity function, $m^*_R = 24.48 \pm 0.15$.
\markcite{Shapley:01}Shapley {et~al.} (2001), based on near-infrared
imaging of $z \sim 3$ LBGs, finds the corresponding $m^*_K = 20.70
\pm 0.25$, or $M_V^* = -22.21 \pm 0.25 + 5 \log h$ (for $\Omega_m =
0.3, \Omega_\Lambda = 0.7$).  An $R \approx 29.5$ LBG at $z \sim 3$ is
thus $\approx 10^{8.8} L_\odot$ and has a $\approx 10^{7.3} M_\odot$
central black hole according to the \markcite{Magorrian:98}Magorrian
{et~al.} (1998) black hole mass -- galaxy luminosity relation.  Modulo
uncertainties in black hole mass evolution, galaxy density profile
evolution, and evolution in the mass-to-light ratio, these numbers imply
a central blackhole well within the $10^{6-8} M_\odot$ black hole mass
range required to produce tidal disruption events \markcite{Rees:88}(Rees
1988), as well as a galaxy luminosity well within the plateau range in the
\markcite{Magorrian:99}Magorrian \& Tremaine (1999) tidal disruption rate.

Correcting the \markcite{Magorrian:99}Magorrian \& Tremaine (1999) tidal disruption rate for the $1 +
z$ time dilation, assuming that the resultant flares would only be visible
for $\approx 2$ months, and using the number of highly-lensed $z \sim
3$ LBGs determined above, we estimate an $\approx 4 \times 10^{-7}$
probability that \pals1\ is associated with the gravitationally-lensed
optical flare caused by a tidal disruption event at $z \approx 3.3$.

\section{Conclusions}

We report the discovery of a transient source, ``\pals1'', in the field
of Abell~267.  At discovery, the source was undetected in the UV, $U_n
> 24.7$ (2$\sigma$), but had blue colors from 5000 \AA\ to 9000 \AA,
with $m_{\rm AB} \approx 22.2$ at 6500 \AA\ and a blue spectral index,
$\beta_\nu = 1.0 \pm 0.4$.  The spectral energy distribution suggested
\pals1\ to be an ideal candidate lensed Lyman-break galaxy at $z \sim
3$; indeed, our lensing model for Abell~267 places \pals1\ near a cusp
in the source plane for this redshift.  However, unlike Lyman-break
galaxies, subsequent observations showed ``\pals1'' to fade by over
four magnitudes in seven months.  We discuss potential interpretations
of \pals1.  Similar sources likely contaminate Lyman-break surveys
\markcite{Steidel:99}(\eg Steidel {et~al.} 1999) at some level.

First we consider the possibility that \pals1\ is a SN, either at the
cluster redshift ($z = 0.23$) or in a slightly background ($z = 0.278$)
spiral galaxy 4\farcs5 WNW of \pals1.  We briefly note that finding
gravitationally-lensed type~Ia SNe behind galaxy clusters is potentially
quite interesting as a technique to remove the mass-sheet degeneracy
that arises when weak lensing studies of gravitational shear are used
to infer the cluster mass \markcite{Kolatt:98, Holz:01}(\eg Kolatt \& Bartelmann 1998; Holz 2001).  In short,
because the shear field is insensitive to magnification, a uniform sheet
of matter anywhere between the source and the observer will remain
undetected in weak lensing analyses.  Detection of a strongly-lensed
standard candle provides direct measurement of the magnification,
thus lifting the mass-sheet degeneracy.  We show that normal type~I
and type~II SNe at all epochs have colors inconsistent with \pals1.
A somewhat rare SN~1991T-like type~Ia SN has similar colors to \pals1\
near peak brightness for $z = 0.278$, but would be 1.5~magnitudes more
luminous than \pals1.  Known, underluminous type~Ia SNe have inconsistent
spectral energy distributions.  We conclude that \pals1\ is unlikely
to be associated with any known SN, though the possibility exists that
\pals1\ marks the first discovery of a new type of rare SN.

Next, we consider the scenario that the $U$-band dropout signature of
\pals1\ is indeed associated with interstellar and intergalactic
hydrogen absorption for a source at $z \sim 3$.  We argue that the most
likely explanation for both the SED and light curve is tidal disruption
of a star by a massive black hole in the core of a galaxy.
\markcite{Rees:88}Rees (1988) show the remnants of such a star forms a stream of
stellar matter in far-ranging orbits \markcite{Cannizzo:90}(see also Cannizzo, Lee, \& Goodman 1990).
Assuming equal fractions of mass in equal binding energy intervals, the
inferred accretion rate declines as $t^{-5/3}$, consistent with the
light curve of \pals1.  Such events are predicted to be very luminous
and to give rise to extremely blue spectra, corresponding to effective
temperatures $>$ 10,000~K \markcite{Ulmer:99}(\eg Ulmer 1999).  \markcite{Loeb:97}Loeb \& Ulmer (1997)
note that such events potentially provide a crude measure of the
central black hole mass.  In particular, for black holes more massive
than $\approx 10^8 M_\odot$, the tidal radius lies inside the innermost
stable orbit ($6 G M_{\rm bh} / c^2$) and stars are swallowed whole,
without much emission \markcite{Hills:74}(Hills 1974).  Flares produced by
tidally-disrupted stars have the potential to extend studies of the
accretion history of the universe and the relation between black holes
and their host galaxies to quiescent galaxies and to very high
redshift.

It is somewhat dissatisfying that since \pals1\ was a transient
phenomenon, we may never know what it truly was.  A crude estimate
of the relative event rates for $z = 0.23$ peculiar, type~Ia cluster
SNe and $z \sim 3.3$ stellar disruption flares argue that the former
is perhaps a few orders of magnitude more likely, depending upon the
existence and frequency of cluster SNe which would exactly match the
colors and brightness of \pals1.  What observations could help unwrap
this mystery?  Foremost would be archival observations of Abell~267
obtained in Summer/Fall 2001.  Though the $R$-band light curves for
both post-maximum $z \sim 0.25$ SNe and tidal disruption events are
similar (Fig.\ \ref{evo.plot}), the expected color evolutions are quite
different and would likely provide conclusive information (\eg Fig.\
\ref{SNe.plot}).  Moderately-deep $R$-band imaging from the few weeks
prior to the discovery images would also unambiguously distinguish a
pre-maximum SN from a tidally-disrupted stellar flare.  Further study of
Abell~267 would also be useful.  Though challenging, color information
for the $R \approx 28$ presumed host could distinguish a $z \approx
0.25$ star-forming region from a $z \approx 3.3$ LBG.  Deep imaging
could also help improve the mass modeling of Abell~267 and identify
potential counter images.  Finally, improved knowledge about type~Ia
SNe, particularly at ultraviolet wavelengths, will also be useful.
The growing census of well-studied type~Ia SNe will improve our estimate
of the fraction that are peculiar, as well as elucidate the breadth of
this variety.  If indeed an underluminous 1991T-like SN were identified,
it would strengthen the case the \pals1\ is a similar event.

We note that \pals1\ was found with a rather shallow survey of $\approx
100$ galaxy clusters and, assuming it is a tidal disruption flare
at $z \approx 3.3$, the amplification was at most a factor of 10.
This implies that deeper surveys (not restricted to clusters) could
commonly find more disruption events.  Distant field \markcite{Wittman:02}(\eg the Deep
Lens Survey; Wittman {et~al.} 2002) and nearby galaxy \markcite{Li:03b}(\eg the robotic
KAIT program; Li {et~al.} 2003a) surveys should be routinely searched for such
events at galaxy centers.  Future variability surveys (\eg PanSTARRS
and LSST, expecting to reach more than 2.5 magnitudes deeper than PALS)
might also be optimized to discover disruption events.  Finally, we
note that transients like \pals1\ are relevant for future SN search
programs, such as the {\it Supernova Acceleration Probe} ({\it SNAP}).
If \pals1\ had indeed been a disruption event, its existence (well-above
the {\it SNAP} threshold), with a light curve similar to a type~Ia SN,
strengthens the case for well-sampled multi-color light curves with
spectra taken near peak. If \pals1\ had been a peculiar type~Ia SN,
such sources, otherwise assumed to be normal type~Ia SNe, would act as
contaminants for cosmology programs.

\acknowledgements{We gratefully acknowledge Avishay Gal-Yam, Avi Loeb,
and Peter Wannier for useful comments.  We thank Chuck Steidel for use
of his $U_n$-band filter, Graham Smith for the {\em HST} image of Abell
267, and David Kaplan for NIRC observations.  Haakon Dahle and Ashish
Mahabal assisted in obtaining archival data.  The work of DS was
carried out at Jet Propulsion Laboratory, California Institute of
Technology, under a contract with NASA.  PGvD is supported by NASA
through Hubble Fellowship HF-01126.01-99A awarded by the Space
Telescope Science Institute.  DJS would like to acknowledge financial
support from NASA's Graduate Student Research Program, NASA grant No.
NAGT-50449.  JSB acknowledges support by the Fannie and John Hertz
Foundation.  MS acknowledges support from a PPARC fellowship.  TT
acknowledges support from NASA through Hubble Fellowship grant
HF-01167.01.  We also extend our thanks to Martin Kirby, under whose
influence this project commenced.}


\clearpage


\clearpage
\begin{deluxetable}{cccc}
\tablecaption{UT 2001 July 20 photometry of \pals1}
\tablehead{
\colhead{$U_{\rm AB}$} &
\colhead{$g_{\rm AB}$} &
\colhead{$r_{\rm AB}$} &
\colhead{$i_{\rm AB}$}}
\startdata
$>24.7$ & $21.96 \pm 0.12$  & $22.16 \pm 0.16$ & $22.56 \pm 0.12$ \\
\enddata

\tablecomments{$U$-band photometry is a 2$\sigma$ upper limit.
Photometry has been corrected for Galactic extinction of $E(B-V) =
0.025$ towards Abell~267, determined from the dust maps of
\markcite{Schlegel:98}Schlegel, Finkbeiner, \& Davis (1998).}

\label{sed.table}
\end{deluxetable}

\begin{deluxetable}{ccc}
\tablecaption{$R$-band photometry of \pals1}
\tablehead{
\colhead{Julian Date} &
\colhead{$R_{\rm AB}$} &
\colhead{Telescope/Instrument}}
\startdata
46679.3 & $>23.7$ & DPOSS \\
48159.6 & $>22.5$ & DPOSS \\
48511.3 & $>23.7$ & DPOSS \\
48624.3 & $>23.8$ & JKT \\
48915.8 & $>22.6$ & DPOSS \\
49698.3 & $>24.2$ & WHT \\
51045.3 & $>24.8$ & JKT \\
51081.3 & $>23.9$ & SDSS \\
51109.8 & $>26.9$ & UH 2.2m\\
51860.4 & $27.7^{+0.4}_{-0.3}\, ^{\dag}$ & {\em HST} \\
52110.5 & $22.14 \pm 0.14$ & Palomar $200''$ \\
52115.6 & $22.3 \pm 0.3$ & Keck guider\\
52203.4 & $25.30 \pm 0.15$ & Keck/ESI\\
52227.3 & $26.1 \pm 0.3$ & Keck/ESI\\
52321.2 & $>26.4$ & Keck\\
\enddata

\tablecomments{Limits are 2$\sigma$ upper limits. $^{\dag}${\it HST} detection
is presumed associated with the host star-forming region or galaxy.}

\label{tableLC}
\end{deluxetable}

\end{document}